\documentclass[draft]{agujournal2019}
\usepackage{url} 
\usepackage{lineno}
\usepackage[inline]{trackchanges} 
\usepackage{soul}
\usepackage{framed}
\usepackage{amsmath}
\usepackage{amsfonts}
\usepackage{comment}

\usepackage{natbib}

\usepackage[skip=0.333\baselineskip]{caption}
\usepackage{subcaption}
\usepackage{xcolor}

\setstcolor{black} 

\nolinenumbers
\usepackage{amsmath}
\usepackage{mdframed}
\usepackage[none]{hyphenat}
\usepackage{todonotes}

\renewcommand{\st}[1]{\unskip}
\newcommand{\newTxt}{\textcolor{black}}

\draftfalse

\journalname{Geophysical Research Letters}

\begin{document}
%
%


\title{Surface temperature extremes produced by huge machine learning hindcasts of summer 2023}

%
%




\authors{Mark Risser\affil{1,*}, Ankur Mahesh\affil{2,1,*}, William D. Collins\affil{3,1}, Joshua North\affil{1}, Boris Bonev\affil{4}, Karthik Kashinath\affil{4}, Thorsten Kurth\affil{4}, Michael S. Pritchard\affil{4,5}, and Shashank Subramanian\affil{6}
}


\affiliation{1}{Earth and Environmental Sciences, Lawrence Berkeley National Laboratory (LBNL), California, USA}
\affiliation{2}{Earth and Planetary Science, University of California, Berkeley, California, USA}
\affiliation{3}{The School of Environmental Sciences, University of East Anglia, Norwich, UK}
\affiliation{4}{NVIDIA, Santa Clara, USA}
\affiliation{5}{Department of Earth System Science, University of California, Irvine, California, USA}
\affiliation{6}{National Energy Research Scientific Computing Center, LBNL, Berkeley, California, USA}

\affiliation{*}{These authors contributed equally to this work.}




\correspondingauthor{Mark D. Risser}{mdrisser@lbl.gov}



\begin{keypoints}
\item Surface temperature extremes generated by a huge ML-based ensemble exceed both observations and those from a traditional forecast ensemble. 

\item For 30\% of the globe, the ML ensemble predicts more intense heatwaves than those derived from extreme value theory applied to NWP ensembles. 

\item \newTxt{The statistical properties of a huge ensemble differ intrinsically from those of a smaller ensemble ``boosted'' to asymptotic size using GEV.} 

\end{keypoints}

%
%

%
%





\section*{Abstract}
\noindent
\st{The summer of 2023 was the second hottest on record, with numerous extreme heatwaves across the globe.}
\newTxt{The summer of 2023 brought record-breaking heat extremes across the globe. In this work, we investigate how much worse these extremes might have become, given identical large-scale conditions. Numerical weather prediction (NWP) ensembles simulate alternate heatwave storylines, but their limited sample size diminishes their statistical coverage of observed heatwaves and ability to represent worst-case events.} Using the Spherical Fourier Neural Operator machine learning (ML) weather model, we generated a huge ensemble of 7,424 \st{weather scenarios simulating} \newTxt{storyline forecasts of} summer temperature extremes. \st{The ML ensemble produced extreme heatwave scenarios exceeding temperatures from reanalysis and numerical weather prediction ensembles} \newTxt{Despite being constrained by the same initial and boundary conditions and using training data from the same modeling system,} \st{our results show that} \newTxt{the ML ensemble produced extreme heatwave scenarios exceeding temperatures from NWP} ensembles. \st{the ML model's extreme surface temperatures were not unusual for approximately two-thirds of the global land area. However, for the other} 
\newTxt{For 30\% of global land areas,} ML-generated extreme events were \st{well} outside the prediction envelope from extrapolating smaller \newTxt{NWP} ensembles with extreme value theory. 
\st{The ML ensemble also readily generates storyline simulations of humid heat extremes, which yield more dangerous categories of public safety alerts than can be simulated from smaller ensembles.} 
\st{This research highlights the potential of huge ensemble simulations to improve prediction of worst-case humid and dry temperature extremes.}  \newTxt{This research demonstrates that very large ensembles substantially improve upon NWP-based characterization of worst-case heat extremes.}

\section*{Plain Language Summary}
\noindent Devastating heatwaves now threaten vast areas of the Earth's surface, presenting significant challenges for adaptation and impact planning. The extreme, record-breaking heatwaves experienced during the summer of 2023 highlight the necessity of understanding the full plausible range of how hot temperatures can become.
This study uses machine learning (ML) to simulate alternate \newTxt{plausible} heatwave scenarios, with the ability to explore more extreme outcomes than current simulation capabilities.
We find that the ML model generates extreme heatwaves that exceeded observed temperatures for summer 2023 and those predicted by traditional forecasting methods. For nearly one-third of the global land area, including parts of Greenland, Russia, Alaska, and China, the ML-generated heatwaves were well outside the prediction range of traditional forecasts. 
\st{We furthermore show how the ML model yields important information about the risks associated with humid heat extremes.}
Our study highlights the capacity of ML to simulate realistic and otherwise unanticipated extreme weather events at a fraction of computational cost of state-of-the-art weather forecasting tools.
\section{Introduction} \label{sec:intro}

Extreme heatwaves are among the most damaging hazards associated with a changing climate, and understanding how intense such events can plausibly become is an important scientific challenge. The 2021 heatwave in the North American Pacific Northwest and British Columbia is a compelling example of a recent extreme heatwave \citep{White_2023}. It was generated by interactions among atmospheric circulation \citep{Wang_2023}, sea-surface temperature patterns, and soil moisture \citep{Schumacher_2022}. These collectively amplified the event beyond five standard-deviations away from the historical median surface  air temperatures \citep{Bartusek_2022} with anomalies relative to climatology as high as 25$^\circ$C \citep{Emerton_2022}.  The daily maximum temperatures are well above the upper bounds set by Generalized Extreme Value (GEV) theory applied to the climate records for that region \citep{Bercos-Hickey_2022}. Events of this severity can be reproduced using traditional Earth System  and numerical weather prediction models using large ensembles \citep{McKinnon_2022}, plausible counterfactual ``storylines'' identified using ensemble boosting techniques \citep{Fischer_2023}, and ensemble weather and seasonal forecasts \citep{Leach_2024, Finkel2023, Kelder2020}.

\newTxt{Heatwaves like the 2021 Pacific Northwest event underscore the need to simulate and quantify worst-case heatwaves, particularly as the climate warms.}  We assess heatwaves during the boreal summer of 2023, which was the hottest summer in the observational record at the time \citep{WMO2023} and during the last 2000 years \citep{Esper2024,Hegerl2024}. The temperature anomalies during summer 2023 significantly exceeded those predicted by observationally-calibrated statistical climate models \citep{Schmidt2024}, \newTxt{largely due to the effect of internal variability \citep{Gyuleva2025}.}  The annually averaged temperature of 2023 was higher than a climatological baseline during 1850-1900 by 1.43${}^\circ$C  \citep{Forster2024}.  \st{We elected to construct our huge ensemble of hindcasts for this summer to understand how several types of extremes, including heatwaves but also tropical cyclones and atmospheric rivers, would respond to these warmer conditions.} \newTxt{The numerous extreme and record-breaking heatwaves that occurred in the boreal summer of 2023 \citep{zachariah2023extreme} make this specific season an important test case for quantifying worst-case heatwaves with traditional and machine-learning (ML) methods.}


\newTxt{Our primary objective in this work is to quantify how much hotter summer 2023 extremes could plausibly have been. 
From a climatological perspective, such questions are typically answered using statistical \citep{Risser2025data} or theory-based \citep{zhang2023upper,risser2025upper} approaches, but these cannot account for the large-scale dynamical environment responsible for driving the most extreme events (e.g., the PNW heatwave described above). Even techniques such as ensemble boosting \citep{Fischer_2023} cannot isolate a common background weather state and instead explore extremes associated with different initial and boundary conditions, across climatology. To answer this question, we simulate alternate weather trajectories, or storylines \citep{Shepherd2018}, initialized from each day of summer 2023. }

\newTxt{NWP forecast ensembles offer one existing approach, simulating alternate trajectories from the observed atmosphere to quantify how much worse a heatwave could have been. However, their computational cost limits them to 50-100 members, sampling on average only $2\sigma$ from the ensemble mean \citep{Mahesh2024hugeensemblespartii}. In summer 2023, about 6\% of observed land temperatures fell outside the NWP ensemble range (Supplemental Figure S10), and these misses correspond precisely the tail relevant to low-likelihood, high-impact events \citep{Seneviratne2022_book}. Furthermore, NWP ensembles underestimated peak temperatures by several Kelvin on the hottest days (Supplemental Figure S11), failing to capture extremes that actually occurred. ML weather emulators overcome this limitation, generating large, skillful ensembles at low computational cost \citep{Mahesh2024hugeensemblesidesign, Mahesh2024hugeensemblespartii}. A 7,424-member ML ensemble extends the sampled range to approximately $4\sigma$ and misses only 0.6\% of observed events (Supplemental Figure S10). This improved coverage is not achieved by indiscriminately overpredicting heatwaves \citep{Lerch2017}: it matches NWP's forecast skill and spread-error ratio, with improved resolution of the tail of the forecast distribution \citep{Mahesh2024hugeensemblespartii}.}

\newTxt{Our goal is not simply to generate ever-larger ensembles: sample maxima grow with ensemble size regardless of information content. Instead, statistical coverage of observed events provides a principled criterion: since the large ML ensemble reduces the miss rate for summer 2023 heat extremes from 5.8\% to 0.6\%, it serves as a statistically grounded reference for quantifying worst-case heatwaves. At this improved coverage, we ask how much hotter summer 2023 extremes could plausibly have been, beyond what is captured by observations (a single realization) or NWP ensembles (roughly 50 realizations). Using identical boundary and initial conditions and training data from the same parent modeling system, we compare worst-case temperature extremes from large versus small ensembles, the latter boosted to asymptotic size via GEV theory.}

\section{Data and Methods}
\label{sec:DataMethods}

\subsection{\newTxt{Requirements for generating plausible  summer 2023 counterfactuals}} \label{subsec:requirements}

\newTxt{
As discussed in Section~\ref{sec:intro}, our primary objective is to quantify how much worse boreal summertime heat extremes might \textit{plausibly} have become in 2023. To answer this question robustly and precisely, we select data sets and ensembles according to the following criteria:
\begin{enumerate}
\item The initial and boundary conditions must be identical to those that were realized during the boreal summer of 2023.
\item We restrict our attention to short-range hindcasts in order to preserve memory of these initial and boundary conditions and avoid drift toward the generating model's climatology.
\item Candidate models must accurately predict weather events departing from climatology in summer 2023, ensuring the physical realism of simulated weather.
\item Candidate models must exhibit fast growth in ensemble spread, with a spread-to-RMSE ratio of approximately one at the lead times of interest.
\end{enumerate}
The last criterion ensures that the ensemble's growth in spread will encompass both plausibly worse \textit{and} plausibly better counterfactual futures. If the ensemble members are exchangeable with one another and with reality, then the spread-error ratio of the ensemble will be near one \citep{Fortin2014}. This is a necessary condition for the ensemble to represent statistically plausible alternate versions of the atmospheric roll out.
We note that techniques such as ensemble boosting \citep{Fischer_2023} or pseudo-global warming storylines \citep{Shepherd2018} do not satisfy these criteria, as they rely on different initial and boundary conditions and may not display appropriate growth in ensemble spread relative to errors. Violating requirement \#2, statistical approaches such as \cite{Risser2025data} that analyze seasonal extremes over many decades are likewise not applicable to the scientific question here, as they target the climatological rather than meteorological characteristics of heatwaves.
}


\subsection{\st{Overview of HENS} \newTxt{Huge ML-based weather forecast ensembles (HENS)}}
\label{ssec:HENS}

We created a  huge ensemble (HENS) of weather extremes using an ML emulator of global numerical weather reanalyses and simulations of the recent climate record.  
\st{Our goal is to conduct the first studies of whether massive ensembles from ML emulators improve the ability to simulate ``low likelihood, high-impact'' (LLHI) weather events, using terminology from the sixth Intergovernmental Panel on Climate Change Report, which states that we currently have low confidence in current and future projections of LLHIs (Seneviratne et al., 2021).
Studies of LLHI extreme weather and climate events require massive ensembles to capture long tails of multivariate distributions (Hall et al., 2023). Until recently, generating 1,000- or 10,000-member ensembles of high-resolution hindcasts was exceptionally challenging because of prohibitive compute costs. The scientific community has recognized, however, that such massive ensembles are needed to characterize LLHIs (Thompson et al., 2017).} 
HENS is an ensemble hindcast consisting of 7,424 members per initialization day \citep{Mahesh2024hugeensemblesidesign, Mahesh2024hugeensemblespartii}. \newTxt{The ensemble is initialized every day from 01~June 2023 to 31~August 2023, thus meeting criteria \#1 in Section~\ref{subsec:requirements}. To meet criteria \#2, each member is a short 15-day simulation that represents a different way the weather could have evolved, based on the actual initial conditions from 2023.  } The ensemble is generated using the Spherical Fourier Neural Operator (SFNO), a global machine-learning weather forecasting model \citep{Bonev2023}. To produce the 7,424 ensemble members for a given initialization, we begin with initial conditions from ERA5 and introduce perturbations using the bred vector method \newTxt{with 29 independently trained SFNO models} \citep{Mahesh2024hugeensemblesidesign}. 
\st{This method generates perturbations aligned with the fastest-growing directions in the model state space, producing a diverse set of initial conditions that reflect uncertainties in the atmospheric state.  We combine bred vector initial condition perturbations with an estimate of model uncertainty by using 29 independently trained SFNO models.} 
HENS includes 12 meteorological variables on a 0.25-degree horizontal grid  with 6-hourly temporal resolution (see Table S1 in the Supporting Information) \citep{Mahesh2024hugeensemblespartii}. In this study, however, we focus on 2m temperature (t2m). 
\newTxt{Previously, SFNO was evaluated using globally and temporally averaged metrics \citep{Bonev2023}, which underweight rare extremes. To address criterion \#3, we compute the Extreme Forecast Index (EFI), a climatology-relative measure of forecast anomalies; SFNO's EFI matches that of the physics-based IFS \citep[IFS;][]{ECMWF}. Our ML ensemble also matches physics-based ensembles in reliability and discrimination of extremes \citep{Mahesh2024hugeensemblesidesign}.  Compared with a smaller ensemble, HENS significantly improves the representation of the warm tail of the forecast distribution, as measured by a 17\% improvement in the outcome-weighted version of the Continuous Ranked Probability Score \citep{Mahesh2024hugeensemblespartii}. Although SFNO, like other deterministic AI models, produces blurred, coarser-resolution forecasts than ERA5 and IFS, its spectra remain stable during rollout \citep[unlike, e.g.,][]{Kochkov2024, Lang2024} preserving realistic atmospheric structure at 10-day lead times. This provides an extremes-focused validation pipeline beyond prior work \citep{Bonev2023}. For criterion \#4, we show that the ensemble's spread-error ratio is near one, especially at 10 days \citep{Mahesh2024hugeensemblesidesign}, confirming adequate ensemble spread relative to error.}
\newTxt{Collectively, these diagnostics show HENS is realistic and fit-for-purpose for simulating tail events. In summer 2023, a physics-based NWP model (IFS) missed true observed high temperatures at 10 times the rate of HENS (Supplemental Figure S10), despite the fact that HENS and IFS have comparable forecast skill \citep{Mahesh2024hugeensemblespartii}. This improved coverage reflects real events: across high latitudes, eastern China, northern Africa, and Central America, IFS underestimated observed maximum temperatures by several Kelvin, whereas HENS captured these extremes within its envelope (Supplemental Figure S11).}


\newTxt{For comparison with HENS, we also evaluate a 58-member subsample of the huge ensemble, denoted HENS-58, in order to explicitly test the ML ensemble against NWP ensembles of comparable size. 
The huge ensemble is generated using initial condition perturbations and multiple trained models, and these 58 members represent a minimal set of the ensemble by choosing all trained checkpoints and two perturbations per member.  This smaller ensemble was validated in \citet{Mahesh2024hugeensemblesidesign}. }

\subsection{Comparison data sets} \label{subsec:comparisondata}

For verification, we use the ERA5 global reanalysis prepared by the European Centre for Medium Range Forecasting \citep[ECMWF;][]{Hersbach2020}.  \st{This reanalysis provides all the data required to construct the daily maximum surface temperatures.}  ERA5 is available at hourly frequencies on a rectilinear latitude/longitude grid with a spatial resolution of 0.25${}^\circ$.   As HENS is trained against ERA5 (from 1979-2016, with summer 2023 withheld from the training dataset), comparison of these two datasets quantifies the additional information content of the HENS ensemble relative to the ERA5 baseline. 

To benchmark the information content of existing NWP ensembles, we use the 50-member ECMWF Integrated Forecast System \citep[IFS;][]{ECMWF}. IFS consists of medium-range 15-day integrations from day 0 to day 15, \newTxt{which also meet the criteria for inclusion in our study described in Section~\ref{subsec:requirements}. More details on the IFS data are provided in Section 1 of the Supporting Information.} 
\st{with an effective resolution of 18~km version CY47R3,234
used from June 1 to June 27, 2023; ECMWF, 2021) or 9~km (version CY48R1, used from June 28 to August 31, 2023; ECMWF, 2023). The switch in IFS versions was dictated by ECMWF's schedule of updates  (ECMWF, 2025a).  The integrations are initialized twice daily at 00 and 12 UTC. In the vertical, the integrations have 91 (CY47R3) or 137 (CY48R1) levels, with the model top at 1 Pa.  The two-dimensional IFS fields we use in this study were interpolated to 0.25-degree resolution by ECMWF before downloading.  As ERA5 is based upon a recent but earlier cycle of the IFS system CY41R2; Hersbach et al., 2020), comparison of HENS against IFS quantifies the additional information content of the larger HENS ensemble size relative to that of IFS along with small systematic offsets between CY41R2 and CY47R3/CY48R1.}
\newTxt{We use IFS as a comparison benchmark in this work because it is widely recognized for its high forecast quality.  It provides a current state-of-the-science approach to simulate alternate possible realizations of summer 2023 with physics-based models. It is important to note, however, that HENS is trained to mimic ERA5 rather than IFS, and agreement with IFS plays no role in the HENS training criteria. Our previous work demonstrates that HENS and IFS achieve comparable forecast skill \citep{Mahesh2024hugeensemblesidesign}, though this does not guarantee similarity in the extreme states generated by each ensemble. Notable structural differences exist between the two systems: in its current form, HENS receives no explicit information about the land surface state \citep[all the input variables to the model are atmospheric variables; see][]{Mahesh2024hugeensemblesidesign}, whereas IFS incorporates a fully coupled land surface model \citep{ECMWF}. Despite these differences, a detailed comparison of the two systems offers a valuable assessment of ML-based ensemble forecasting in the context of the most extreme heatwave events.}

\subsection{Selection of extreme measurements} \label{sec:goals_data}

We extract the summertime maximum 2-meter air temperature and heat index values from HENS, the IFS ensemble, and ERA5. For each of the 92 days in the summer of 2023 (01 June 2023 to 31 August 2023), IFS provides a 50-member ensemble of a 15-day forecast at a six-hourly resolution (61 total time steps, with forecast lead times of $0, 6, \dots, 360$ hours), for each $0.25^\circ \times 0.25^\circ$ longitude-latitude grid box over the globe.  HENS provides an analogous ensemble with 7,424 members \newTxt{(or, in the case of the subset ensemble, 58 members)}. 
At each IFS grid box, we extract the maximum t2m \st{and heat index} measurement for each ensemble member from the $\{240, 246, 252, 258\}$ hour (i.e., 10-day) lead times for all 92 days. 
The output of this procedure is a set of 50 extreme six-hourly measurements at each $0.25^\circ \times 0.25^\circ$ grid box for 2m temperature. \st{and the heat index.} We similarly extract extreme measurements at each grid box for the HENS ensemble\newTxt{s} and ERA5. Since our primary interest is in extreme temperatures over land, we filter out all grid boxes with a land fraction of less than 0.75. 

\subsection{Statistical methods and probabilistic framing}

In this work we assess whether extreme value theory applied to a conventional ${\cal{O}}(50)$-member ensemble can extrapolate into the tails of the forecast distribution as effectively as direct sampling with a huge ensemble.  The main statistical analysis is therefore fitting generalized extreme value (GEV) distributions to measurements of 2m temperature \st{and the heat index} from the IFS \newTxt{and HENS-58} ensembles over the summer of 2023. We use Bayesian methods to estimate the GEV distributions of the \st{IFS} \newTxt{smaller} ensembles as well as conduct uncertainty quantification. A detailed description of our methods is described in Section 2 of the Supporting Information. \newTxt{The main objective in fitting extreme value distributions to small ensembles is to extrapolate beyond the range of the data to arbitrarily large quantiles, effectively converting a small sample to an ``infinite'' sample.}

\newTxt{
The validity of any statistical model for extreme measurements depends on whether its underlying assumptions are satisfied in practice. We verify this for GEV distribution applied to the IFS and HENS-58 data in two ways. First, we argue that the ensembles, which span on the order of 50 maxima across 4 six-hourly time steps $\times$ 92 days = 368 data points, are sufficient for the asymptotic GEV behavior to emerge, particularly since extreme temperatures are bounded \citep{dombry2019maximum} as shown in Supplemental Figures S1 and S2. Second, we verify goodness of fit using Q-Q plots, which compare empirical quantiles of the data against theoretical quantiles of the fitted distribution.
Points falling along the 45$^\circ$ (1-1) line (and within the statistical uncertainty) are evidence of a good fit.
Results presented in Section 2.3 of the Supporting Information show strong agreement between empirical and theoretical quantiles for 2m temperatures in both ensembles, providing compelling evidence that the GEV assumptions are satisfied. We can therefore be confident that the fitted distributions reliably characterize the upper tail behavior of extreme temperatures and we can safely extrapolate beyond the support of the data.}

\section{Results} 
\label{sec:Results}

As discussed in Section~\ref{sec:DataMethods}, we address our key objective 
using two modeling systems (ECMWF's numerical models, underlying ERA5 and IFS, and the ML model SFNO underlying HENS) at varying ensemble sizes: single realizations (ERA5), 50-member ensembles (IFS, HENS-58), and very large or effectively infinite ensembles (HENS, IFS-GEV, HENS-58-GEV). To separate effects of modeling system from ensemble size, we first compare across sizes within each system (Section~\ref{subsec:withinMod}), then across systems at comparable sizes (Section~\ref{subsec:btwnMod}). For the large and infinite ensembles, we use the 99.9th percentile as our extreme threshold, rather than the GEV upper bound or ensemble maximum, since this threshold contains the ERA5 t2m maxima for nearly 98\% of global land area (with diminishing returns at higher percentiles; Table S2, Supporting Information). Using a percentile also has the advantage of being well-defined: unlike the ensemble maximum, which grows unbounded with ensemble size, a percentile estimate converges with decreasing uncertainty as ensemble size increases.

\subsection{Comparing within models across ensemble size} \label{subsec:withinMod}

\begin{figure*}[!t]
\centering
\includegraphics[width = \textwidth]{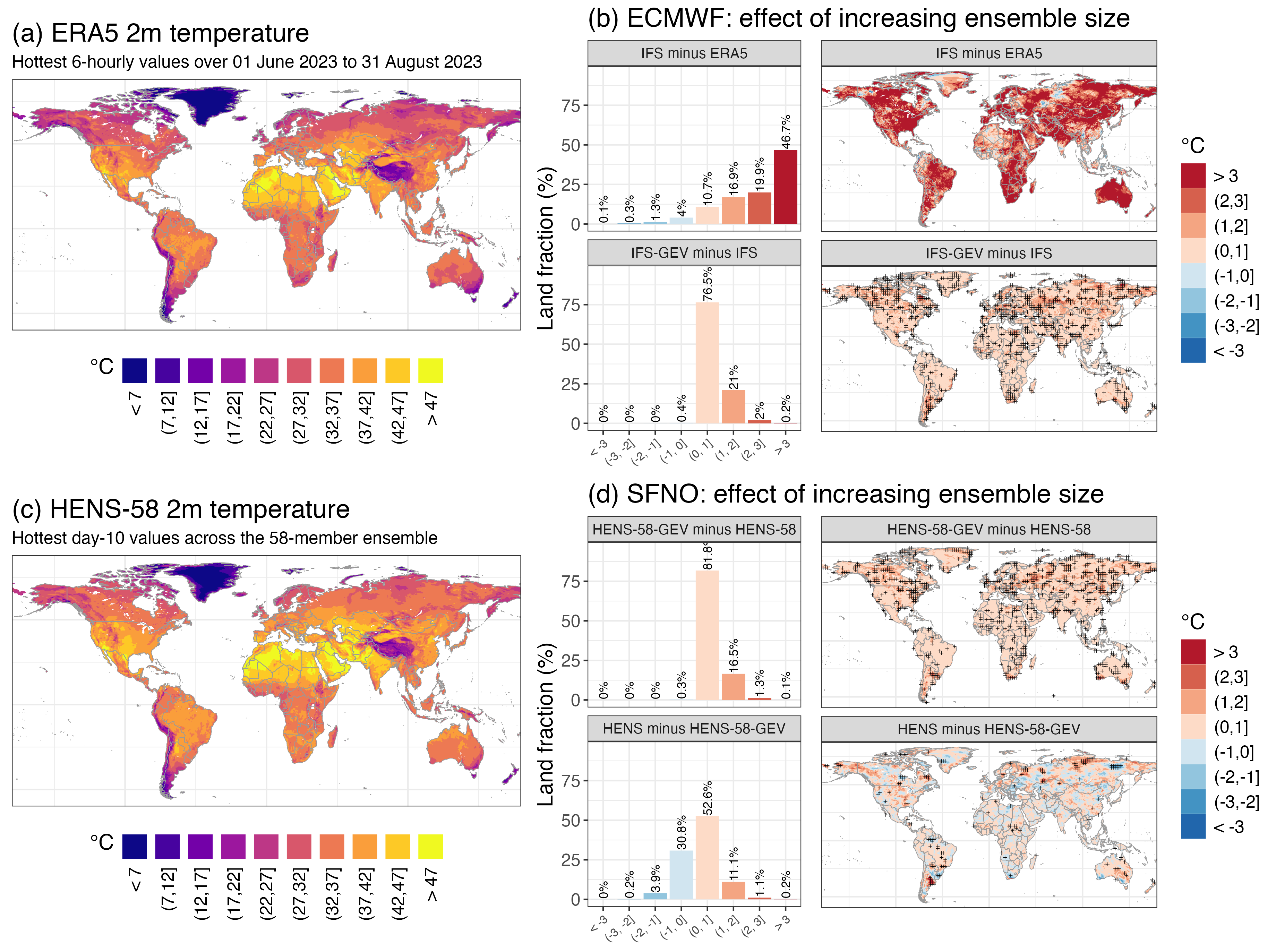}
\caption{\newTxt{
\textbf{Within-model comparisons across ensemble size.} Panels (a) and (b) examine the effect of increasing ensemble size for the ECMWF modeling family, comparing the realized 2m temperatures from ERA5 (a single realization) with IFS (maximum of a 50-member ensemble) with IFS-GEV (extrapolated to the 99.9th percentile). Panels (c) and (d) consider Spherical Fourier Neural Operators, comparing the maximum from the 58-member subset ensemble with HENS-58-GEV (extrapolated to the 99.9th percentile) with HENS (the 99.9th percentile of the huge ensemble). Stippling indicates that the 95\% uncertainty interval for the difference \textit{does not} include zero.}}
\label{fig:compwithin}
\end{figure*}

Figure~\ref{fig:compwithin} compares extreme 2m temperatures across ensemble sizes within each modeling system. We begin with the ECMWF system, contrasting the realized surface temperature maxima from ERA5 (Figure~\ref{fig:compwithin}a) with those from IFS and IFS-GEV (Figure~\ref{fig:compwithin}b). This comparison does not involve the machine learning emulator, but establishes the state of the science prior to our generation of HENS. The ERA5 data have the hottest temperatures ($53.6^\circ$C) in the Kerman Province of Iran, near $(58.5^\circ$E, $30^\circ$N$)$; otherwise, only limited areas in Northern Africa, the Arabian Peninsula, Iran, and Central Asia experienced temperatures exceeding $47^\circ$C.


The IFS maxima exceed ERA5 for 94.2\% of the land area (Figure~\ref{fig:compwithin}b). In many cases this exceedance is more than $3^\circ$C and exceeds $15^\circ$C in isolated areas. Comparing the IFS 50-member ensemble maxima to the IFS-GEV extreme threshold, we expect IFS-GEV to always be larger, since it represents an extrapolation to a much higher quantile ($0.999$) relative to the ensemble maximum (which should roughly correspond to the $1 - 1/50 = 0.98$ quantile). This expectation is verified: except for a very small area in Central Asia, the IFS-GEV threshold is warmer across all global land areas considered. For approximately three-fourths of the globe this exceedance is modest (within $1^\circ$C), though in some areas the IFS-GEV estimates exceed the IFS maxima by more than $2^\circ$C. Notably, the IFS-GEV threshold is significantly larger than the IFS maxima for only a fraction of the globe (stippling in Figure~\ref{fig:compwithin}b indicates statistical significance), because the uncertainty in the $0.999$ quantile estimated from only 50 data points is quite large: the 95\% uncertainty interval is approximately $3^\circ$C wide on average, and in some cases exceeds $10^\circ$C.

We next compare SFNO ensembles, contrasting temperature extremes from the HENS-58 ensemble (Figure~\ref{fig:compwithin}c) with those from HENS-58-GEV and HENS (Figure~\ref{fig:compwithin}d). 
As with the IFS vs. IFS-GEV comparison above, we expect HENS-58-GEV to exceed the HENS-58 maxima everywhere, since it represents an extrapolation to a higher quantile. This expectation is again verified: the HENS-58-GEV extreme threshold estimates are larger than the HENS-58 maxima across all land regions except a small area in eastern China. As before, many of these differences are not statistically significant, owing to the large width of the HENS-58-GEV uncertainty intervals (on average $2.7^\circ$C and as large as $9.7^\circ$C; see Supplemental Figure S5).

Finally, we compare HENS-58-GEV to HENS directly. Having verified that the GEV assumptions are satisfied for the HENS-58 data, our null hypothesis is that the HENS-58-GEV estimates of the $0.999$ quantile should be unbiased and the uncertainty intervals appropriately calibrated. Neither expectation is fully borne out. The HENS-58-GEV estimates are, on average, nearly $0.25^\circ$C too cool relative to the HENS threshold. Additionally, while the distribution of differences is roughly symmetric and centered near zero, there are non-negligible fractions of the globe where the HENS threshold is more than $1^\circ$C warmer (in some cases $4^\circ$C warmer) than its HENS-58-GEV counterpart. Collectively, this indicates that HENS-58-GEV based on this specific subset underestimates the magnitude of extreme 2m temperatures in these regions compared to the larger HENS ensemble. Although the majority of these differences are not statistically significant, this is largely a consequence of the wide 95\% uncertainty intervals (Supplemental Figure S5b). Furthermore, the uncertainty intervals are slightly overconfident: the 90\% credible intervals from HENS-58-GEV cover the HENS quantity for only 85.5\% of grid boxes, and the 95\% credible intervals cover only 91.7\% of the globe. This pattern of under-coverage persists across all uncertainty levels (Supplemental Figure S5a.).

In summary, comparisons within each modeling system reveal a consistent picture: larger ensembles and GEV-based extrapolations systematically identify more extreme temperature thresholds than smaller ensembles or reanalysis alone, as expected. The 50-member IFS ensemble exceeds ERA5 over 94.2\% of the land area, and IFS-GEV further extends these extremes. For SFNO, HENS-58-GEV exhibits a modest cool bias and slight overconfidence in its uncertainty intervals relative to the very large HENS ensemble. Furthermore, the HENS-58-GEV uncertainty intervals are quite wide, which challenges their usefulness in practice. Ultimately, these results highlight the limitations of extreme value theory as a substitute for direct, large-ensemble sampling when estimating temperature extremes.

\subsection{Comparing between models for fixed ensemble size} \label{subsec:btwnMod}

\begin{figure*}[!t]
\centering
\includegraphics[width = \textwidth]{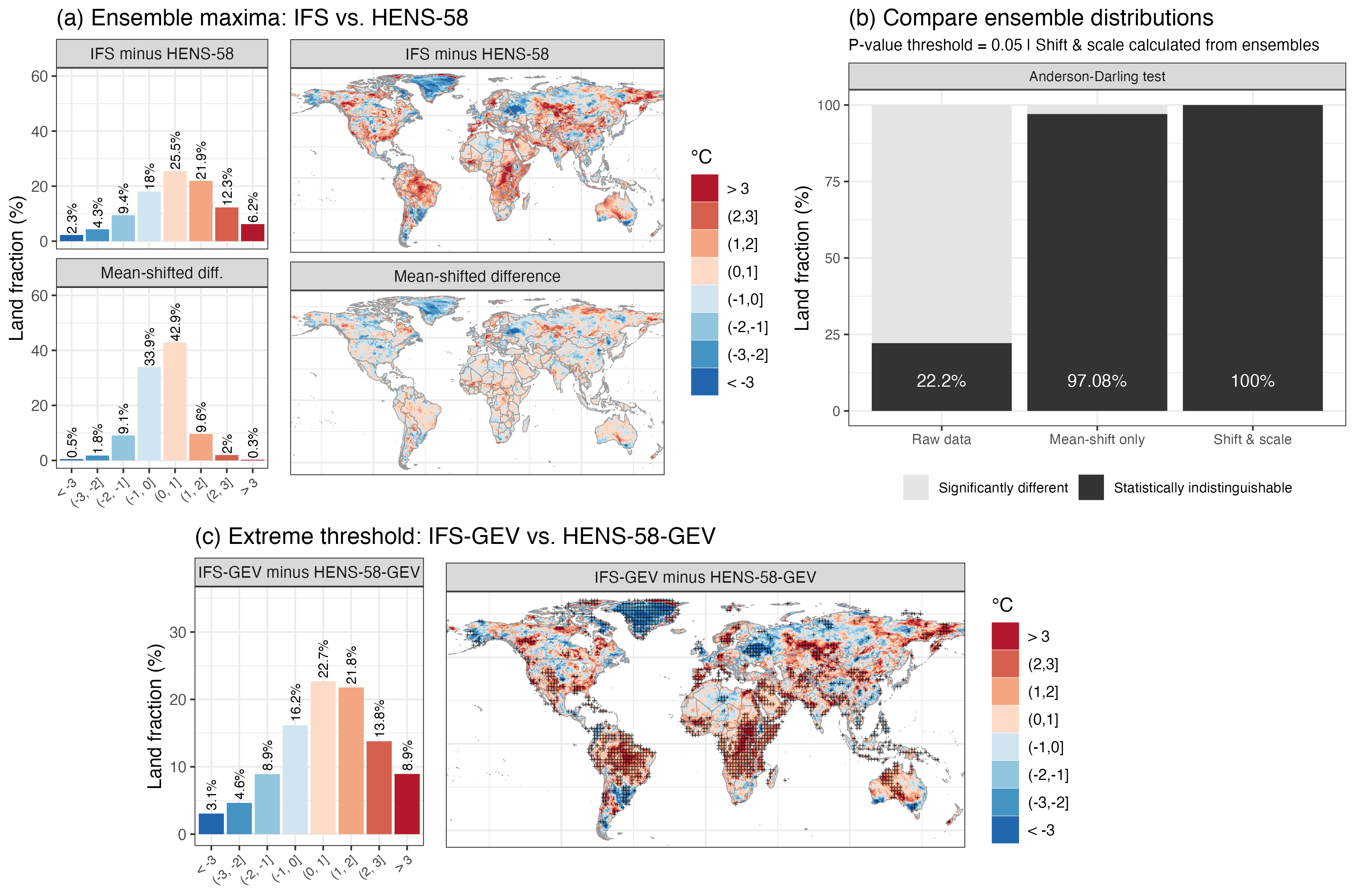}
\caption{\newTxt{\textbf{Between-model comparisons for a given ensemble size.} Panel (a) compares ensemble maxima from IFS and HENS-58 for both the original and mean-shifted data. Panel (b) tallies the fraction of the globe for which the ensemble distributions are statistically indistinguishable according to the Anderson-Darling test for both raw and shifted and rescaled data. Panel (c) considers the extreme threshold (the 99.9th percentile) from IFS-GEV and HENS-58-GEV. In panel (c), stippling indicates that the 95\% uncertainty interval for the difference \textit{does not} include zero.}}
\label{fig:compbtwn}
\end{figure*}



We now shift our focus to between-model comparisons while holding ensemble size constant. Figure~\ref{fig:compbtwn} compares the ensemble maxima from IFS and HENS-58 (each with approximately 50 members) as well as the inferred IFS-GEV and HENS-58-GEV distributions. The distribution of differences in ensemble maxima between IFS and HENS-58 (top row of Figure~\ref{fig:compbtwn}a) contains many non-zero values with clearly structured spatial patterns, and differences exceed $2^\circ$C in absolute value for more than 50\% of the globe. Overall, the IFS maxima are warmer than the HENS-58 maxima, though it is notable that the ML emulator produces temperatures at least $2^\circ$C warmer than the NWP ensemble for approximately 6.6\% of the land areas considered. These differences extend across the entire ensemble distribution: for the raw data, the two ensembles are statistically indistinguishable for only 22.2\% of the globe (Figure~\ref{fig:compbtwn}b). However, nearly all of these differences can be attributed to a mean shift: removing the ensemble mean at each grid box from both the ensemble maxima (bottom row of Figure~\ref{fig:compbtwn}a) and each ensemble member (Figure~\ref{fig:compbtwn}b) substantially reduces the magnitude of differences, leaving the two ensembles statistically indistinguishable over nearly the entire globe (97.1\%). A further rescaling by the grid-box standard deviation reconciles the remaining 3\%.  Further research is needed to determine the systematic differences in the ensemble mean summertime maximum estimates between IFS and HENS.


As with the empirical ensembles, the IFS-GEV and HENS-58-GEV also yield statistically different estimates of the 99.9th percentile storyline.  The IFS-GEV exceeds the HENS-58-GEV threshold by more than $2^\circ$C for a large fraction of the globe, yet in other (albeit more limited) regions the HENS-58-GEV threshold is more than $2^\circ$C warmer. The largest absolute differences are statistically significant (indicated by hatching in Figure~\ref{fig:compbtwn}c), accounting for approximately 40\% of the globe. A comparison of GEV distributions fit to both ensembles reveals that the discrepancy between the two systems is primarily a location shift (Supplemental Figure S4; Section 2.4 of the Supporting Information), and the spatial pattern of differences in GEV location parameters closely matches the spatial pattern of differences in ensemble maxima shown in Figure~\ref{fig:compbtwn}(b) (pattern correlation $= 0.79$). Importantly, GEV shape parameter estimates from IFS and HENS-58 are statistically indistinguishable (Supplemental Figure S4), suggesting that the ML emulator may perform comparably to IFS in characterizing the extreme upper tail of 2m temperatures.

\begin{figure*}[!t]
\centering
\includegraphics[width = \textwidth]
{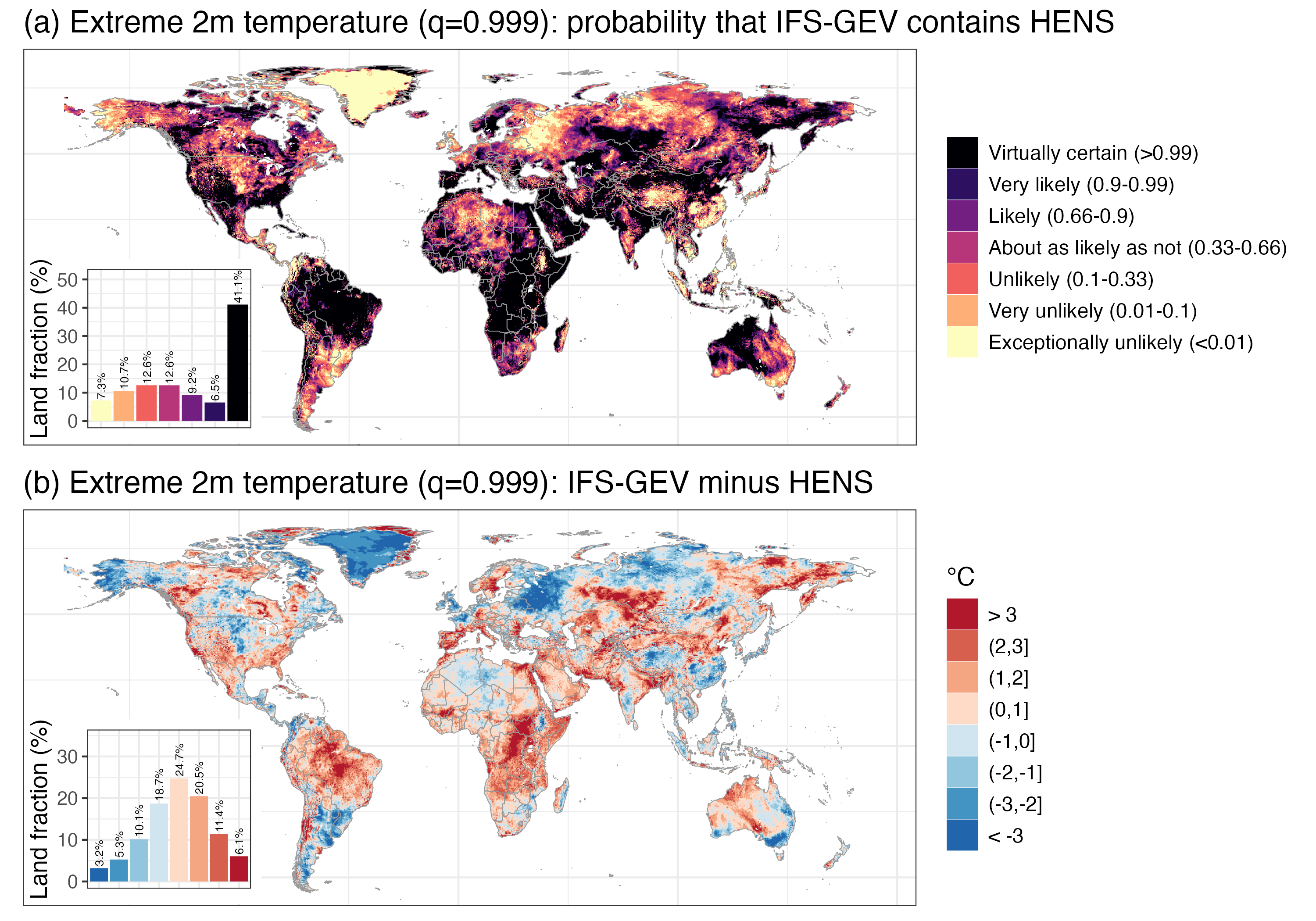}
\caption{\newTxt{Comparison of GEV extreme event thresholds for 2m temperature ($99.9\%$ percentile) from IFS with corresponding empirical estimates calculated from the full HENS ensemble. Panel (a) tallies the probability that the IFS-GEV threshold contains (is larger than) the HENS threshold; lighter colors indicate that HENS values are more extreme than the IFS-GEV estimates. Panel (b) tallies the difference in GEV event threshold (posterior median) and the HENS estimate. Panel insets show the land fraction in each color category.}}
\label{fig:probcontain}
\end{figure*}

Finally, we compare the full HENS ensemble with IFS-GEV thresholds for the 99.9th percentile of 2m temperatures. To compare a single value (HENS) with a statistical estimate (IFS-GEV), we calculate the posterior probability that the HENS threshold is contained within the IFS-GEV threshold, converting these to IPCC-inspired confidence categories \citep{mastrandrea2010guidance,Risser2025data} (Figure~\ref{fig:probcontain}a). We also compute differences between the IFS-GEV posterior median and the HENS threshold. Since IFS-GEV extrapolates to an effectively infinite ensemble while HENS samples a very large ensemble, our null hypothesis is that the two should be indistinguishable, based on: (1) identical initial and boundary conditions; (2) similar effective ensemble sizes; (3) spread-error ratios near 1 at this lead time; and (4) ERA5 (HENS's training data) being partly derived from IFS short-range forecasts. This null hypothesis is violated by the data.

In Figure~\ref{fig:probcontain}(a), lighter colors indicate places where HENS values are more extreme than expected from IFS-GEV alone. We are \textit{virtually certain} that IFS-GEV contains the HENS threshold for over one-third of land area, with another third \textit{about as likely as not} contained. For the remaining 30.6\%, HENS thresholds are definitively more extreme; for 7.3\%, they are \textit{exceptionally unlikely} relative to IFS-GEV extrapolation---concentrated in the northern high latitudes (most of Greenland, eastern/northern Russia, Alaska) but also eastern China, northern Africa, and parts of central North America. Where HENS exceeds IFS-GEV (blue in Figure~\ref{fig:probcontain}b), the exceedance is typically $1$--$3^\circ$C, occasionally larger. Observed temperature records confirm HENS generates realistic high-latitude temperatures, e.g., in Greenland (Supporting Information, Section~3). These results are insensitive to the specific extreme threshold used (Supplemental Figure S7).


In summary, between-model comparisons reveal systematic but largely mean-shift-driven differences between IFS and HENS-58. 
Critically, GEV shape parameter estimates from IFS and HENS-58 are statistically indistinguishable, suggesting comparable characterization of the extreme upper tail of 2m temperatures. However, comparing the full HENS ensemble against IFS-GEV thresholds shows HENS generating \textit{exceptionally unlikely} temperatures over 7.3\% of global land area.
This indicates that HENS identifies plausible worst-case temperatures exceeding IFS-GEV extrapolation, and reaffirms the value of brute-force ensemble sampling over GEV extrapolation alone. This conclusion is further verified by comparing the HENS empirical large ensemble to a GEV extrapolation of HENS-58 (Supplemental Figure S6). 

\color{black}

\subsection{\newTxt{Overall conclusions from within- and between-model comparisons}} \label{subsec:overall}

\newTxt{
The chain of comparisons from ERA5 to HENS (Sections~\ref{subsec:withinMod} and \ref{subsec:btwnMod}) isolates the contributions of ensemble size, statistical extrapolation, and model type to changes in estimated extreme temperatures. From ERA5 to IFS (same model, larger ensemble), IFS exceeds ERA5 for 94.2\% of land area. From IFS to IFS-GEV (statistical extrapolation to a larger effective ensemble), increases are widespread but modest, within $1^\circ$C for roughly three-fourths of the globe. From IFS-GEV to HENS-58-GEV (same extrapolation, different model), systematic differences emerge primarily from a mean shift between IFS and HENS-58, though the two agree well in tail structure, with statistically indistinguishable GEV shape parameters. Finally, from HENS-58-GEV to HENS (same model, brute-force sampling vs.\ extrapolation), GEV-based extrapolation persistently underestimates extreme temperatures: HENS exceeds HENS-58-GEV for 65\% of the globe, and is (at least) \textit{very unlikely} relative to HENS-58-GEV for over 20\% (Supplemental Figure S6a). The largest jumps occur at the first link (ERA5 to IFS) and the last (HENS-58-GEV to HENS), highlighting the dominant roles of ensemble size and the limits of statistical extrapolation for characterizing the most extreme heatwaves.
}

\section{Discussion} \label{sec:Discussion}

In this work, we show that a 7,424-member huge ensemble captures the long upper tails of weather hazards that traditional 50-member NWP ensembles miss. With this petabyte-scale ensemble, we characterize plausible extreme weather events that could have occurred during summer 2023, the hottest summer on record at the time. \st{As summers reaching 2023 levels become increasingly likely due to climate change,} \newTxt{Since the 2023 record temperatures have already been exceeded by subsequent summers,} the ability to fully characterize the envelope of possible low-likelihood, high impact heat extremes will only grow in importance.

A key emergent property of HENS is that its simulated extremes often exceed corresponding predictions from the IFS and HENS-58 GEV distributions. For IFS, the global land surface partitions into three roughly equal regimes: regions where HENS maxima are less than, comparable to, and significantly exceed the GEV thresholds.
This three-way partition is not prescribed by theory and represents an emergent property of HENS. \newTxt{Relative to HENS-58-GEV, these three regimes cover roughly 20\%, 30\%, and 50\% of global land area, respectively (Supplemental Figure~6a). 
}
Ultimately, HENS reveals an envelope of plausible meteorological extremes that traditional extreme value theory and smaller ensembles fail to capture.

\newTxt{
Our analysis in Sections~\ref{subsec:withinMod} and \ref{subsec:btwnMod} yields a puzzling result: IFS is warmer than HENS-58 for 65.9\% of the globe (Figure~\ref{fig:compbtwn}a., top row) and likewise IFS-GEV is warmer than HENS-58-GEV for 67.2\% of the globe (Figure~\ref{fig:compbtwn}c.), yet IFS-GEV contains the HENS threshold with \textit{virtual certainty} for only 41.1\% of the globe (Figure~\ref{fig:probcontain}a.). The resolution to this apparent paradox lies in what GEV extrapolation of HENS-58 fails to capture relative to HENS: as shown in Figure~\ref{fig:compwithin}(d), HENS is warmer than HENS-58-GEV for 65\% of the globe. While the HENS-58 data satisfy GEV assumptions and provide reasonable estimates of the ``true'' extreme thresholds (as measured by HENS), the associated uncertainties are large and there are persistent cases where HENS-58-GEV underestimates the most extreme temperatures. Ultimately, this illustrates a fundamental limitation of GEV-based extrapolation for thresholds well beyond the range of the available sample size.
When computational resources permit, our results therefore highlight the value of brute-force sampling, which makes no reliance on statistical modeling assumptions.
}

\newTxt{We restrict our analysis to the direct effect of ensemble size, comparing huge and small ensembles of short, conditioned forecasts at the same lead time. This differs from comparing ML ensembles to \textit{all-time} climatology, and assessing whether they can generate events more extreme than their climatological training data \citep{Sun2025,Zhang2026}. We intentionally condition 10-day hindcasts on the summer 2023 atmospheric state, since our goal is to produce storylines of that season rather than climatological return periods. Using huge ensembles to estimate return intervals---a forefront of ML weather prediction---would require a different design, with more initial dates and longer rollouts to ensure independent samples.  Complementary methods address these challenges, including rare-event sampling beyond climatology \citep{Lancelin2025}, improving skill for record-breaking extremes \citep{Zhang2026}, and conformal postprocessing for distribution-free coverage \citep{Gopakumar2026, Asch2026}; these improve the forecasting system itself and can be pursued alongside larger ensembles.}

\newTxt{We also restrict our analysis to a single ML emulator, carefully designed and validated to be fit-for-purpose for this science question. Exploring other models is nontrivial: they are generally less validated, lack a readily available ensemble analogue, and in any case share SFNO's characteristics (same training data, same spatial and temporal resolution). Future huge ensembles should use ML models trained end-to-end as ensemble forecasts, evaluated with the extreme diagnostics pipeline of \cite{Mahesh2024hugeensemblesidesign} before such experiments are conducted.}

More broadly, this work establishes a new approach for conducting storyline simulations \st{targeting novel classes} of extremes and compound hazards. 
\st{A current major barrier to storyline simulations is that existing methods demand significant programmer expertise and computational resources, limiting their application in regions that lack access to both.} Machine learning weather models offer a turnkey solution, enabling global storyline simulations without the infrastructure and specialized knowledge required by conventional approaches.
\newTxt{We acknowledge that computational and resource limitations may prevent the generation and analysis of huge ensembles at each 0.25-degree global grid cell. However, ML-based forecasting remains a broadly accessible tool for specific regions and smaller ensembles.} \st{For example,} A single 15-day ML forecast \st{from the ML model} takes approximately one minute, which is orders of magnitude faster than comparable numerical simulations \citep{Mahesh2024hugeensemblesidesign},  \st{Furthermore,} \newTxt{and} forecasts can be easily initialized from publicly available reanalysis data sets such as ERA5. This democratization of tools and models for constructing storyline simulations presents an important opportunity for their broad and systematic application for extreme events in all corners of the globe.

\section*{Open Research Section}

{
The HENS code, datasets, and models are all stored at \url{https://doi.org/10.5061/dryad.2rbnzs80n}, via Data Dryad \citep{hens_dataset}. The code is integrated with Zenodo at the prior DOI. We include the code to train SFNO, conduct ensemble inference with bred vectors and multiple checkpoints, and scoring and analysis code. We also open-source the model weights of the trained SFNO. See the README of the DOI for information on how to use the codebase and for the permissive license associated with the code and data. The code is available via the Lawrence Berkeley National Lab-modified BSD license, and the data is available with a CC0 license. 
}



{
Data from the Programme for Monitoring of the Greenland Ice Sheet (PROMICE) are provided by the Geological Survey of Denmark and Greenland (GEUS) at \url{http://www.promice.dk} \citep{promice_data}. \texttt{ZAC}, \texttt{LYN}, \texttt{FRE} and \texttt{NUK\_K} stations are financially supported by the Glaciobasis programme as part of the Greenland Ecosystem Monitoring project (\url{https://g-e-m.dk/}). The \texttt{NUK\_K} station is owned and maintained by Asiaq Greenland Survey. The WEG stations are funded and maintained by Jakob Abermann at the Department of Geography and Regional Science of the University of Graz. The \texttt{RED\_L} station is funded and maintained by Rainer Prinz at the Department of Atmospheric and Cryospheric Sciences of the University of Innsbruck. The \texttt{NUK\_B} station is funded and maintained by James Lea at the University of Liverpool. The \texttt{SER\_B} station is funded and maintained by Anders Bjørk at the Department of Geosciences and Natural Resource Management of the University of Copenhagen.
}

\acknowledgments

{This research was supported by the Director, Office of Science, Office of Biological and Environmental Research of the U.S. Department of Energy under Contract No. DE-AC02-05CH11231 and by the Regional and Global Model Analysis Program area within the Earth and Environmental Systems Modeling Program (MDR, JSN, WDC, AM). The research used resources of the National Energy Research Scientific Computing Center (NERSC), also supported by the Office of Science of the U.S. Department of Energy, under Contract No. DE-AC02-05CH11231. The computation for this paper was supported in part by the DOE Advanced Scientific Computing Research (ASCR) Leadership Computing Challenge (ALCC) 2023-2024 award ‘Huge Ensembles of Weather Extremes using the Fourier Forecasting Neural Network’ to William Collins (LBNL) and the 2024-2025 award ‘Huge Ensembles of Weather Extremes using the Fourier Forecasting Neural Network’ to William Collins (LBNL).}

%
%
\nocite{Coles2001,ECMWF81272,ECMWF81371,ECMWFTimeline, ECMWF,Fausto2021,Hersbach2020,How2022data,Owen1995,Zhang2024}
\bibliography{agusample}

\end{document}